\documentclass[pra,twocolumn,superscriptaddress,showpacs]{revtex4}

\usepackage{amsbsy}
\usepackage{amsfonts}
\usepackage{amssymb}
\usepackage{amsmath}
\usepackage{color}
\usepackage{graphicx}
\usepackage{float}
\usepackage[caption = false]{subfig}

\begin{document}


\title{Laser from a Manybody Correlated Medium}

\author{Eduardo Mascarenhas}
\affiliation{Laboratory of Theoretical Physics of Nanosystems, Ecole Polytechnique F\'{e}d\'{e}rale de Lausanne (EPFL), CH-1015 Lausanne, Switzerland}

\author{Dario Gerace}
\affiliation{Dipartimento di Fisica, Universit\`{a} di Pavia, via Bassi 6, I-27100 Pavia, Italy}

\author{Hugo Flayac}
\affiliation{Laboratory of Theoretical Physics of Nanosystems, Ecole Polytechnique F\'{e}d\'{e}rale de Lausanne (EPFL), CH-1015 Lausanne, Switzerland}

\author{Marcelo F. Santos}
\affiliation{Departamento de F\'{i}sica, Universidade Federal de Minas Gerais, CP 702, 30123-970 Belo Horizonte, Brazil}

\author{Alexia Auff\`{e}ves}
\affiliation{CNRS and Universit\`{e} Grenoble Alpes, Institut N\'{e}el, F-38042, Grenoble, France}

\author{Vincenzo Savona}
\affiliation{Laboratory of Theoretical Physics of Nanosystems, Ecole Polytechnique F\'{e}d\'{e}rale de Lausanne (EPFL), CH-1015 Lausanne, Switzerland}

\begin{abstract}

We consider a non-equilibrium system of interacting emitters described by the XXZ model, whose excitonic transitions are spatially and spectrally coupled to a single mode cavity. We demonstrate that the output radiation field is sensitive to an interplay between the hopping ($J$) and the interactions ($U$) of the excitons. Moderate values of the short-ranged interaction are shown to induce laser with maximal output at the Heisenberg point ($U=J$). In the laser regime, charge-charge correlations emerge and they are shown to strongly depend on the interaction-hopping ratio. In particular, the system shows charge-density correlations below the Heisenberg point and ferromagnetic correlations beyond the Heisenberg point. This contrast to the equilibrium behavior of the XXZ chain occurs since the laser explores highly excited states of the emitters.

\end{abstract}


\maketitle

\section{Introduction} 

The quantum theory of conventional lasers addresses the stimulated emission of radiation from an active medium made of independent and non-interacting emitters coupled to a single resonator mode~\cite{Qnoise,MeW}. Such description is accurate for most gaining media used in standard lasers. However, substantial technological progress has allowed for the microscopic tailoring of manybody correlations of artificial chains of atoms in different architectures, from superconducting~\cite{Koch,Majer,dicarlo2010nat,Pop} to semiconducting technologies \cite{warburton_review,Robledo}. Such progress may provide the natural playground to investigate the impact of quantum interactions on the lasing properties and, more generally, to address the physics of non-equilibrium manybody effects on the statistics of the radiation emitted from more complex and structured active media.

The manybody physics of driven-dissipative, or non-equilibrium, quantum systems has recently attracted substantial interest~\cite{Diehl,Pizorn,Prosen,Sieberer,Kessler,Fermi,Rossini,Valle,Tomadin1,Tomadin2}, envisioning the possibility to observe emergent phenomena that cannot be properly described by classical thermodynamics at equilibrium, such as, e.g., dissipative phase transitions. On a parallel route, the physics of few emitters coupled to a single-mode resonator has been thoroughly investigated in the past few years~\cite{Wiersig2009,Woggon,Holland1,Holland2,laussy2011prb,Auffeves,CoopNos,pagel2015pra,Leymann}, mainly concerning the role of the cavity-mediated correlations between the independent emitters, and discarding the role of direct manybody couplings of the active medium on the emitted radiation properties.

Here we address the rich physics of a peculiar manybody system made of strongly correlated quantum emitters whose elementary excitations are radiatively coupled to a single-mode cavity. We show that a laser regime results from the interplay between kinetic energy (hopping) and electromagnetic interactions (Coulomb repulsion) between the emitters excitations. From a theoretical point of view, we treat such a system through an exact solution of the equations of motion, and the simulations of a stochastic Schr\"{o}dinger equation. We show that the optimal condition for lasing corresponds to the Heisenberg Hamiltonian, in which hopping and interactions have the same strength, leading to a fully symmetric spin model. We show that in this case the set of states that are symmetric with respect to permutation of any two spins become exact eigenstates of the XXZ chain. Such bright-states differ in total magnetization and are resonantly coupled by the light-matter interaction. Therefore, the cavity is efficiently fed with excitations while the system transitions between this subset of XXZ eigenstates.

In the laser regime, the chain of emitters shows charge-density-wave (CDW) order at short range for interactions below the Heisenberg point and long range uniform correlations at the Heisenberg point.
Above the Heisenberg point, the correlations do not present any CDW ordering, in contrast to the equilibrium ground state behavior of the XXZ chain that presents CDW order for large interactions. These results represent another step to put manybody couplings in the context of a quantum laser scenario. We suggest operational and measurable quantities where the manybody nature of the active medium naturally emerges through probing of the coherent radiation emitted from the lasing cavity. 
A natural implementation of the model analyzed in this work can be envisioned for multiple quantum dots coupled to a semiconductor resonator~\cite{strauf06,Lyasota,Albert}, in particular in view of the recent success in fabricating site-controlled quantum dots ~\cite{Lyasota,Schneider_APL_2009,Pelucchi_NapPhot_2013}. Direct Coulomb-mediated coupling between semiconductor quantum dots in particular has been demonstrated in vertically aligned self-assembled quantum dots~\cite{Bayer,Ardelt}, and in distant quantum dots through coupling to an extended Coulomb complex~\cite{SavonaEx}. Alternatively, the coupling of a set of interacting quantum emitters with the mode of an optical cavity is currently realized in ultra-cold atoms in optical lattices~\cite{Greif,Landig}, and designed in circuit-QED~\cite{Kurcz,Quijandria}.

\section{The model and methods}

\begin{figure}
{\includegraphics[width = 3in]{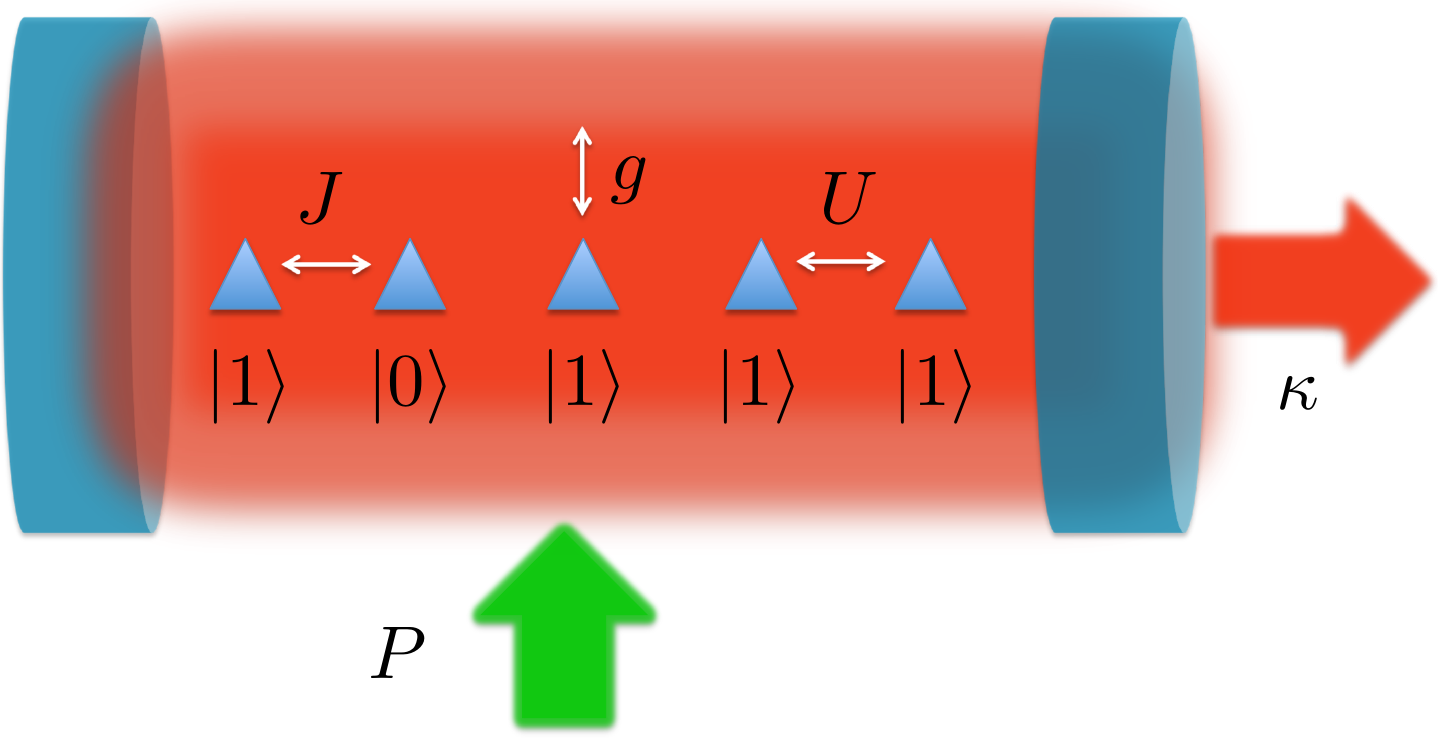}}\\ 
\caption{
Pictorial representation of the laser with a strongly interacting manybody system of emitters as the active medium. 
The microscopic energetic contributions of each possible interacting process are explicitly indicated.
}
\label{SXquema}
\end{figure}

We consider the active medium to be described by a linear chain of degenerate, interacting two-level emitters, modeled by the XXZ spin Hamiltonian
\begin{equation}
H_{\mathrm{XXZ}}=J\sum_{\langle\langle i,j\rangle\rangle}\left[ X_iX_j+Y_iY_j \right]+U\sum_{\langle\langle i,j\rangle\rangle}Z_iZ_j  \, ,
\label{ModelXXZ}
\end{equation}
where $X_i$, $Y_i$ and $Z_i$ are the Pauli matrices, with $\sigma^{\dagger}_i= (X_i + iY_i )/2$ being the raising operator of a single excitation for the $i$-th site. The excitations are allowed to hop from site to site, and $J$ is the hopping strength between nearest neighbor sites. 
Interactions are included in the model via the term proportional to the coupling energy $U$, which models the repulsive or attractive interaction between two neighboring excitations. The resonant interaction with an extra degree of freedom, i.e. the single-mode cavity photon field, is modeled via a Tavis-Cummings Hamiltionian \cite{TavisCummings} 
 \begin{equation}
 H_{\mathrm{TC}}=g\sum_i\left( a\sigma_i^{\dagger}+a^{\dagger}\sigma_i \right) \, ,
 \end{equation}
in which $a^{\dagger}$ is the creation operator for the single-mode bosonic field in the cavity, and $g$ is the light-matter coupling rate between the single emitter excitation at site $i$ and the cavity photons, assumed to be equal for all the emitters. 

In contrast to the commonly studied situation in equilibrium manybody physics, we assume the on-site excitations to be incoherently driven, and the cavity mode to be subject to losses. The dynamics of the system is therefore modeled through the Lindblad-Von-Neumann equation for the density matrix
\begin{equation}
\dot{\rho}=\mathcal{L}(\rho)=-i[H_{\mathrm{XXZ}}+H_{\mathrm{TC}},\rho]+P\sum_i\mathcal{D}_{\sigma^{\dagger}_i}(\rho)+\kappa\mathcal{D}_a(\rho) \, ,
\label{ME}\end{equation}
where the nonunitary part of the dynamics is described by the Lindblad super-operators  
\begin{equation}
\mathcal{D}_x(\rho)=-\frac{1}{2}\left[x^{\dagger}x\rho+\rho x^{\dagger}x\right]+x\rho x^{\dagger} \, .
\end{equation}
Eq. (\ref{ME}) thus models two non-unitary processes: the incoherent driving of the quantum emitters at rate $P$, and the dissipation of the cavity mode at rate $\kappa$. In Fig~(\ref{SXquema}) we illustrate all possible coupling processes, and the relative energy contributions for each single state or pair of states. We finally remark that the present model assumes the resonant condition between the cavity mode and the emitters to hold, thus enabling to express all equations in the frame rotating with the resonant frequency. 

In this work we will focus on the stationary state of Eq.~(\ref{ME}), referred to as the non-equilibrium steady state (NESS). The NESS obeys the linear equation $\mathcal{L}{\rho}_{\mathrm{NESS}}=0$. For the numerical implementation of the steady state solution, it is convenient to map the density matrix representation onto an equivalent vector form. We apply a vectorization procedure, where the density matrix $\rho$ is reshaped into a column vector, here denoted by $|\rho \rangle\rangle$, by concatenating all its columns. 
The markovian Liouvillian is always of the form $\mathcal{L}(\rho)=\sum_iX_i\rho Y_i$ and in oder to vectorize this equation we rely on the property $|\mathcal{L}(\rho) \rangle\rangle=\sum_iY_i^{T}\otimes X_i|\rho \rangle\rangle$, where $X$ and $Y$ are matrices. 
In this vectorized representation we thus define the Liouvillian matrix $\mathcal{L}=\sum_iY_i^{T}\otimes X_i$.
We solve the linear problem $\mathcal{L}|\rho_{\mathrm{NESS}}\rangle\rangle=0$ imposing the trace condition $\langle\langle\openone |\rho \rangle\rangle=\mathrm{tr}\{ \rho\}=1$. We define $\tilde{\mathcal{L}}=\mathcal{L}+|0\rangle\rangle\langle\langle \openone|$, such that $|\rho_{\mathrm{NESS}} \rangle\rangle=\tilde{\mathcal{L}}^{-1}|0\rangle\rangle$, which is the fastest approach to reach the solution, albeit being memory consuming. We have also simulated stochastic Schr\"{o}dinger equations (SSE)~\cite{Qnoise}, that have a much lower memory requirement which allows for addressing bigger systems but are much slower in obtaining the NESS to a given accuracy. 

There are infinitely many SSE corresponding to the same master equation. We have found in practice that with a diffusive equation driven by Wiener noise the observables show lower variance over different runs as compared to the SSE driven by Poisson noise (also known as quantum jumps)~\cite{Jacobs}. Therefore, the diffusive SSE allows for simulating the NESS with fewer realizations. However, we have observed that in the jump version the number of cavity Fock states which have non null probability can always be limited to the number of emitters plus one $(L+1)$ by an appropriate choice of the initial state. Therefore, we use the quantum jump version dynamically tracking the optimal Fock states for representing exactly each trajectory, which allows for addressing very high photon numbers with very few Fock states.

\begin{figure}
{\includegraphics[width = 8.5cm]{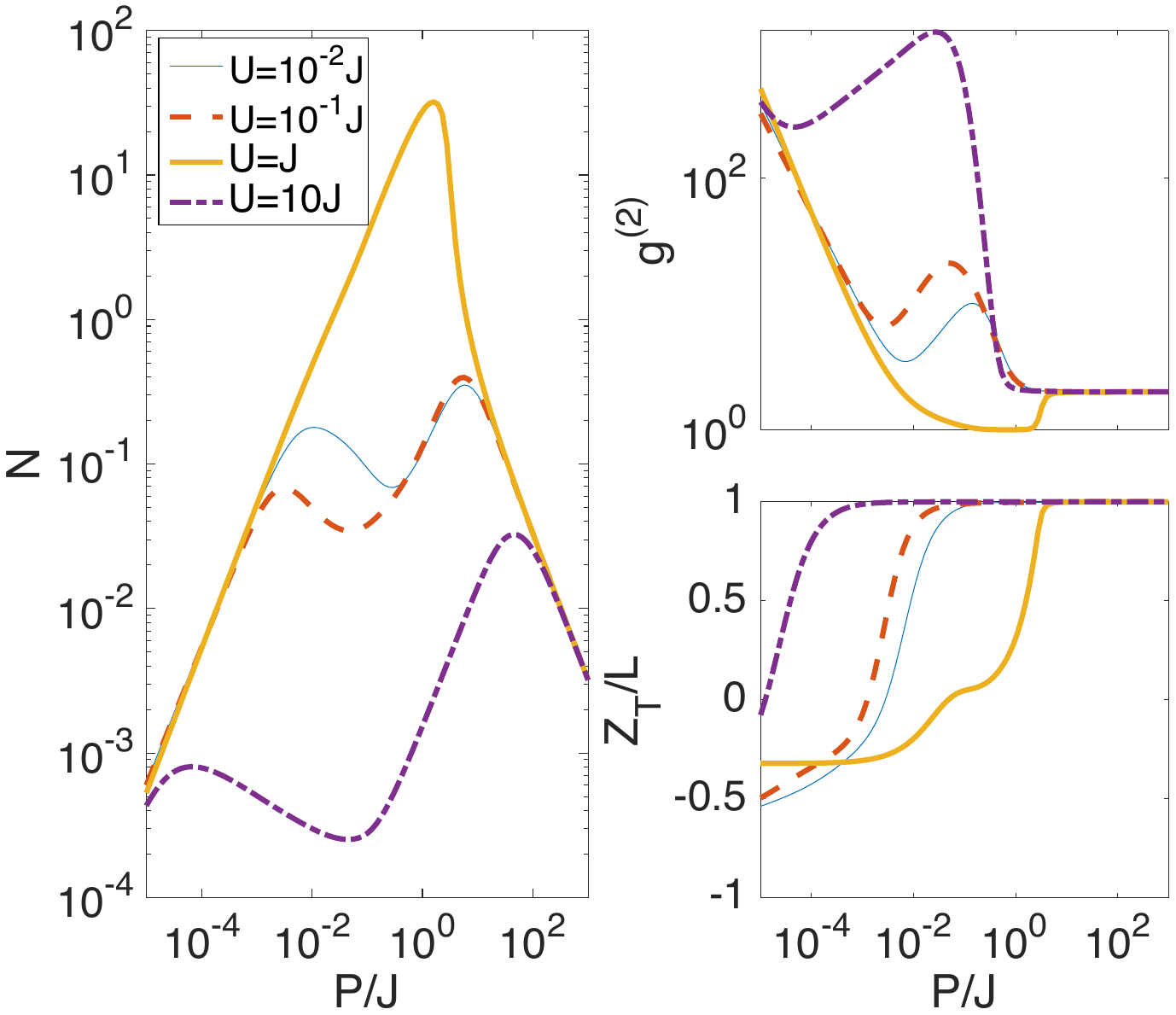}}\\ 
\caption{
Cavity photon number (left panel), second order correlation at zero time delay of the emitted radiation (top right panel), and the $Z$-component of the total atomic spin ($Z_T$) divided by the number of atoms, for the case with $L = 4$ emitters (bottom right panel), as a function of the external pump and for different values of the interaction energy. 
The model parameters assumed in the calculations are $g = 0.1J$ and $\kappa = 0.5g$, respectively.
}
\label{Micro}
\end{figure}

\section{Results}

\subsection{Phenomenology}

In the regime with a small number of emitters, we have an overall number of excitations that allows for an exact solution of the full quantum master equation. We assume that the manybody Hamiltonian is non-negligible, thus we take the hopping rate, $J$, to be larger than the light-matter coupling rate, $g$. At the same time, we choose a high quality factor cavity, such that the dissipation rate $\kappa$ is smaller than the light-matter coupling, which is the typical condition to achieve the lasing regime \cite{Auffeves}. 
We allow for the external incoherent pump rate $P$ to vary, as it is the case for a conventional laser, and we study the cavity output as a function of the short range interaction strength, $U$. 
The results are shown in Fig.~(\ref{Micro}), where we consider a system of four atomic sites in the good cavity regime. 
We immediately notice that for small values of $U/J$, the cavity field accumulates approximately one excitation at most, as it is evident from the plot of the average number of intracavity photons (proportional to the emitted intensity), $N=\langle a^{{\dagger}}a\rangle$. 
As a further figure of merit, we consider the cavity field second-order auto-correlation at zero time delay  
\begin{equation}
g^{(2)}(0)=\frac{\langle a^{\dagger}a^{\dagger}aa\rangle }{\langle a^{\dagger}a\rangle^2} \, , 
\end{equation}
whose value is always above 2 in correspondence with the low cavity emission, i.e. the output cavity field is always bunched. 
Similarly, we see that for $U/J\gg1$ the cavity accumulates even less photons, and the emitted radiation is still bunched. We point out that $g^{(2)}$ always assumes the thermal value of 2 for very strong pump. This can be understood by eliminating the atoms in the regime of strong pump. In this regime the atoms are forced to the maximally magnetized state with all spins pointing up such that they are weakly coupled to the cavity. Thus by making a markovian approximation we may trace over the atomic degrees of freedom deriving a Lindblad term whose sole effect is to pump the cavity with jump operator $a^{\dagger}$. The balance between the cavity dissipation and this effective incoherent pump leads to a thermal state with very low photon number. 
 
\begin{figure}
{\includegraphics[width = 8.5cm]{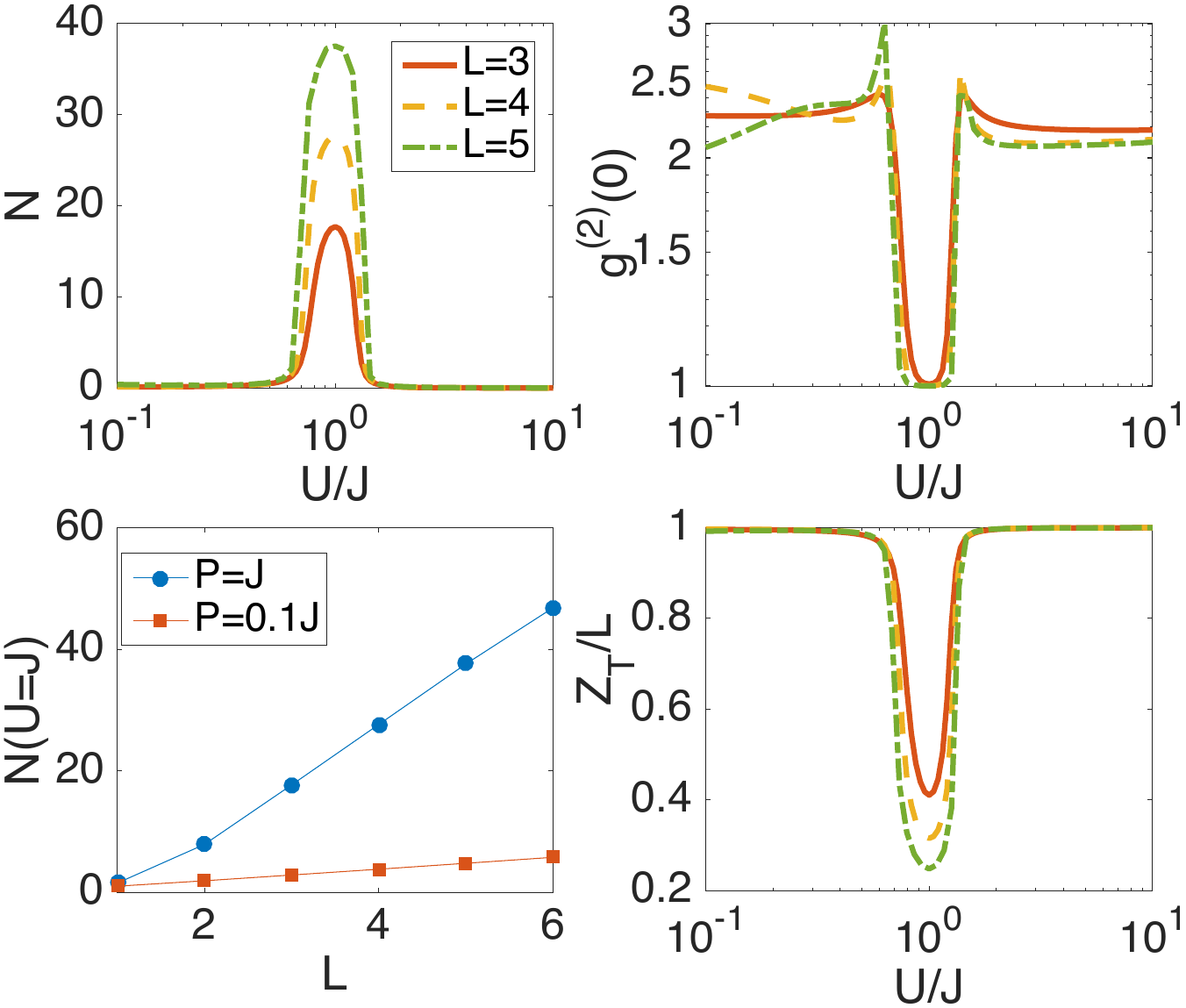}}\\ 
\caption{  
Cavity photon number for $P = J$ as a function of the interaction strength (top left panel), and scaling of the cavity photon number at the Heisenberg point, $U = J$ (bottom left panel). Second-order correlation at zero time delay for the output cavity radiation (right top panel), and $Z$-component of the total atomic spin ($Z_T$) divided by the number of atoms (right bottom panel), as a function of the correlation strength and for different numbers of emitters, in both cases for $P = J$. The system parameters assumed in the calculations are $g = 0.1J$ and $\kappa = 0.5g$, as before.
}
\label{Scaling}
\end{figure}

On the other hand, lasing signatures clearly manifest at intermediate ratios, $U/J\sim 1$. In fact, when the interaction strength is comparable to the hopping rate there is a strong accumulation of photons in the cavity, indicating the occurrence of stimulated emission. Simultaneously, the $g^{(2)}(0)$ reaches the value 1, which is the fingerprint of uncorrelated photon output that is typical of the lasing regime. 
It is also important to analyze the behavior of the total magnetization $Z_T=\sum_iZ_i$ , which shows how the spins react once the stimulated emission sets in. 
In Fig.~(\ref{Micro}) we see that, in the lasing regime ($U\approx J$), $Z_T$ shows a plateau that is known to be a clear signature of stimulated emission, since it represents the clamping of the atomic population. In fact, this shows that even though the external incoherent pump is increased, the atomic population remains essentially unaltered. Thus, the external pump is being efficiently and coherently transferred to the cavity field yielding a strong laser emission from the cavity mirrors.

In Fig.~(\ref{Scaling}) we show how the lasing features emerge at the condition $U/J\approx 1$. Since for $U=J$ the XXZ Hamiltonian (\ref{ModelXXZ}) reduces to the Heisenberg model, we name this condition the \textit{Heisenberg point}.  
As it is seen from the Figure, the cavity population peaks at the Heisenberg point, while $g^{(2)}(0)$ drops to 1, and the total Z-component of the atomic spins approaches zero magnetization per spin on increasing the number of spins. These three features confirm that the cavity output is amplified, the light statistics becomes typical of a laser, and stimulated emission becomes more and more pronounced when hopping and interactions are of the same order. We also show how the cavity output scales with the number of emitters, $L$, at the Heisenberg point. We find a linear scaling of the output with the system size whose slope depends on the pumping rate.

\begin{figure}
{\includegraphics[width = 8.5cm]{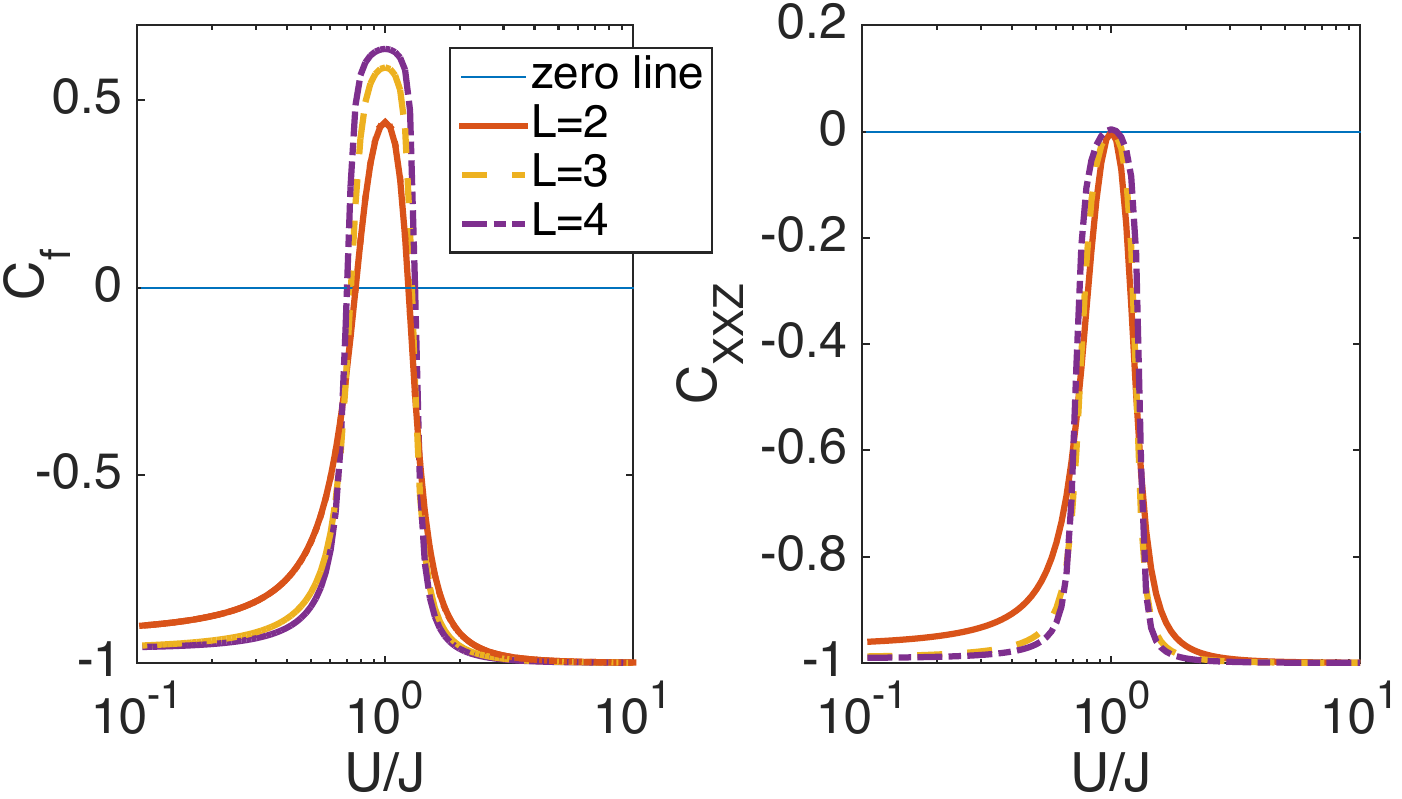}}\\ 
\caption{
Cooperativity of the manybody laser as compared to single emitter-lasers (left panel), as defined in Eq.~({\ref{Cf}}). Cooperativity of the manybody laser as compared to the corresponding standard laser (right panel), as defined in Eq.~({\ref{CXXZ}}). The system parameters used in these simulations are $g = 0.1J$, $\kappa = 0.5g$, and $P=J$.
}
\label{Coop}
\end{figure}

It is worthwhile investigating the impact of the manybody direct couplings of the system on the \textit{cooperativity} of the quantum emitters. A quantitative measure of cooperativity was recently  introduced in Ref.~\onlinecite{CoopNos}. Two different situations are quantitatively compared: the emitters are either coupled to the same cavity mode, or alternatively each of them is coupled to its own cavity mode, at the same resonant energy. The output in the photonic channel, which is proportional to the emitted radiation in each case, is finally compared for these two situations.
The first situation gives rise to an output field (in units of $\kappa$), that is $N(L ,H_{\mathrm{XXZ}})$. In the second case, we measure the sum of the single-cavity outputs from each cavity, where each one contains a single emitter, which is written as $N(1 ,H_{\mathrm{XXZ}}=0)$. 
For a given set of initial conditions, such as pump and dissipation rates, atom-cavity couplings, etc., the system behavior is said to be \textit{cooperative} when the two measurements differ, the difference between them giving direct access to the field that is generated or suppressed by cooperative effects. Then, a cooperativity parameter, or cooperative fraction, can be defined as \cite{CoopNos}
\begin{equation}
C_f= \frac{N(L ,H_{\mathrm{XXZ}})-LN(1)} {N(L ,H_{\mathrm{XXZ}})+LN(1)} \, .
\label{Cf} 
\end{equation}
Such measure is equal to $+1$ when the emitters are maximally and constructively cooperative, while it is equal to $-1$ when they are maximally and destructively cooperative. 
Here, we extend this concept to capture in an isolated manner the impact of the manybody direct couplings of the laser discussed in this paper on the emitters cooperativity.
To this end, we directly compare the manybody laser to the standard laser, by defining the cooperativity parameter of the XXZ Hamiltonian as
\begin{equation}
C_{\mathrm{XXZ}}= \frac{N(L ,H_{\mathrm{XXZ}})-N(L ,H_{\mathrm{XXZ}}=0)} {N(L ,H_{\mathrm{XXZ}})+N(L ,H_{\mathrm{XXZ}}=0)} \, .
\label{CXXZ} 
\end{equation}
These two cooperativity factors defined in Eqs.~(\ref{Cf}-\ref{CXXZ}) are plotted in Fig.~(\ref{Coop}). The results for $C_f$ show that an increasingly sharpened transition from sub- to super-radiance is observed as the number of emitters increases. As expected, the two lasing thresholds are closely located around the maximal cooperativity, which sits at the Heisenberg point. It also becomes clear that both hopping and interactions are detrimental to the laser regime when they act individually, as it is evident from the two regimes in which either one ($U/J\ll1$) or the other ($U/J\gg1$) dominates. 
We notice that the manybody couplings do not allow for extra stimulated emission of radiation, which would result in an even larger intensity of the cavity output.
In fact, the results for $C_{\mathrm{XXZ}}$ show that the manybody laser always emits less radiation than the standard laser, except for the Heisenberg point, in which their emission becomes identical. It is also interesting to stress that, even though the matter-field excitations feature kinetic energy and next-neighbor interactions, these two manybody processes are non-trivially cancelled at $U=J$, and the output intensity becomes the same as the one from a non-interacting laser in which the emitters are not subject to any manybody direct couplings. In the next subsection we describe in detail the mechanism behind the optimal laser regime.

 \begin{figure}
{\includegraphics[width = 6cm]{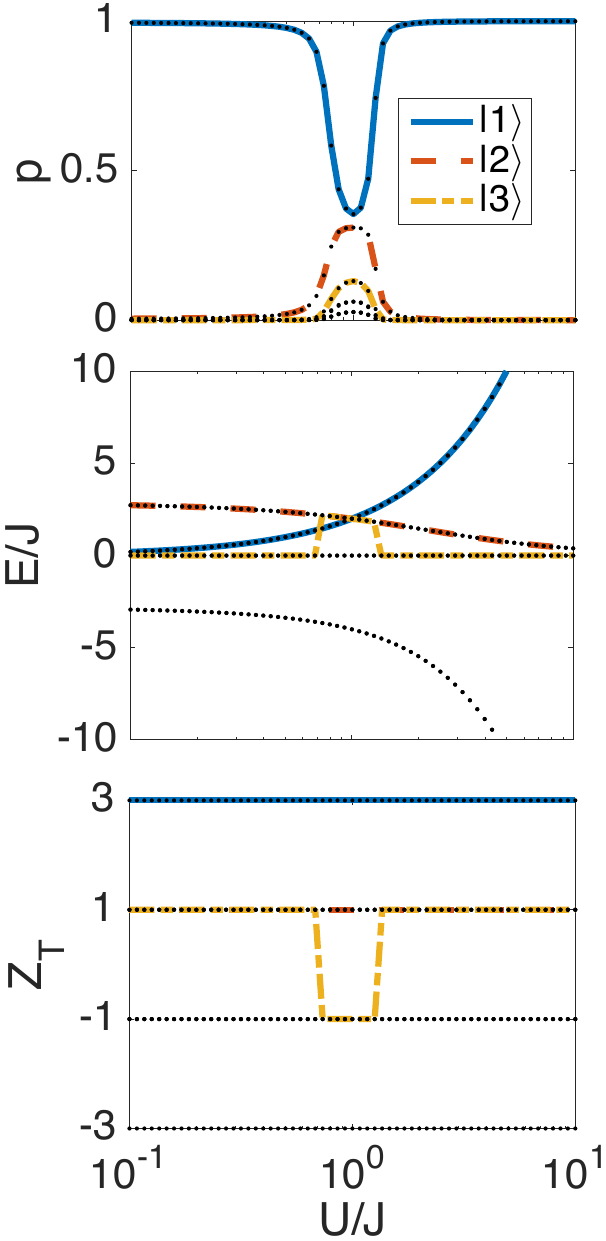}}\\ 
\caption{ 
(Top) The probabilities of the $H_{XXZ}$ eigenstates, as computed from $\rho_{\mathrm{NESS}}$. (Middle) The corresponding eigenenergies. (Bottom) The corresponding magnetization as a function of $U/J$. Data in all three plots were obtained for $L=3$. Highlighted in each plot are the 3 most probable states. The system parameters are $g = 0.1J$, $\kappa = 0.5g$, and $P=J$.
}
\label{EqNonEq3Sites}
\end{figure}

\subsection{The lasing mechanism}

So far, we have mainly focussed on the observables related to the cavity field. However, an analysis in terms of the eigenstates of the spin chain provides insight into the mechanism underlying lasing in the vicinity of the Heisenberg point. 

We recall that, in the case of $L$ independent two-level emitters in a cavity ($H_{\mathrm{XXZ}}=0$), lasing occurs and can be simply explained when the two-level emitters are resonant with the cavity frequency. We denote by $|S,Z_T\rangle$ those, among the eigenstates of the total magnetization, that are even with respect to the exchange of any two spins. These states may be generated by iterating the application of the total spin operator $S=\sum_i \sigma$, starting from the maximally magnetized state $|Z_T=L\rangle=|\uparrow\uparrow\cdots\uparrow\uparrow\rangle$. This iteration determines a sequence of states down to the state $|Z_T=-L\rangle=|\downarrow\downarrow\cdots\downarrow\downarrow\rangle$. The states in this set are then defined (up to a normalization factor) as 
\begin{equation}|S,L-2n\rangle\propto S^n|Z_T=L\rangle, \quad \mathrm{for} \quad n=0,1,\cdots,L.\end{equation}
When all the emitters are in resonance with the cavity, this set of states forms a ladder of equally spaced energy levels with the energy difference matching the cavity frequency. The states $|S,Z_T\rangle$ are also the \emph{bright} states of the spin system, as they are maximally coupled to the cavity mode by the Tavis-Cummings Hamiltonian, which induces transitions among these levels with corresponding absorption/emission of one cavity photon. When operating in the lasing regime, in the limit of large $L$, the balance between pump and dissipation casts the $L$-emitter system into a statistical mixture of states dominated by the zero-magnetization state of the ladder.

In the present case of an interacting spin chain, the optimal lasing is explained by the fact that only at the Heisenberg point $H_{\mathrm{XXZ}}(J=U)=H_{\mathrm{XXX}}$ the set $|S,Z_T=L-2n\rangle$ responsible for the standard laser becomes an exact set of eigenstates of the XXZ Hamiltonian. In fact, these states are fully degenerate in the reference frame of the cavity (i.e. they form a ladder with level spacings coinciding with the cavity resonant frequency), as they obey 
\begin{equation}H_{\mathrm{XXX}}|S,L-2n\rangle=(L-1)J|S,L-2n\rangle\,.\end{equation}
In order to better understand how this set of states enters into the many-body lasing mechanism, we decompose the reduced state of the XXZ chain $\rho_{\mathrm{XXZ}}=\mathrm{tr}_C\{ \rho \}=\sum_{ij}p_{ij}|i\rangle\langle j|$ -- obtained tracing over the cavity degrees of freedom -- in terms of the eigenstates $|i\rangle$ of the bare XXZ Hamiltonian. In Fig.~(\ref{EqNonEq3Sites}) we show the probabilities of each eigenstate $p_i=\langle i|\rho_{\mathrm{XXZ}}|i\rangle$, their eigenenergies, and the corresponding total magnetization computed for a chain of 3 emitters. The three states with highest probabilities are highlighted in color. From this analysis it clearly appears that the chain is completely inverted by the pump, except in the vicinity of the Heisenberg point, where two features arise. First, the most probable states become degenerate in the reference frame of the cavity, as seen in  the middle panel of Fig.~(\ref{EqNonEq3Sites}). Second, the most probable states at the Heisenberg point coincide with the bright states $|S,Z_T\rangle$, namely $|S,3\rangle=|\uparrow\uparrow\uparrow\rangle$, $|S,1\rangle=|\uparrow\uparrow\downarrow\rangle+|\uparrow\downarrow\uparrow\rangle+|\downarrow\uparrow\uparrow\rangle$, and $|S,-1\rangle=|\uparrow\downarrow\downarrow\rangle+|\downarrow\uparrow\downarrow\rangle+|\downarrow\downarrow\uparrow\rangle$. The states thus, together with the state $|S,-3\rangle=|\downarrow\downarrow\downarrow\rangle$ form a ladder in which transitions are allowed by the Tavis-Cummings Hamiltonian. This suggests a clear picture for the onset of lasing. When far from the Heisenberg point, the most probable states in the spin chain are generally out of resonance from the cavity mode, and their corresponding weights are small, thus favoring the fully inverted state induced by the pump. When approaching the condition $J=U$, the states dominating the density matrix form a ladder resonant with the cavity mode. Then lasing sets in and the balance between gain and losses pins the set of emitters at zero magnetization (in the limit of large $L$). The fact that the states in the ladder coincide with the bright states explains why the cooperativity $C_{\mathrm{XXZ}}$ at $J=U$ is exactly zero. 

\begin{figure}
{\includegraphics[width = 8.5cm]{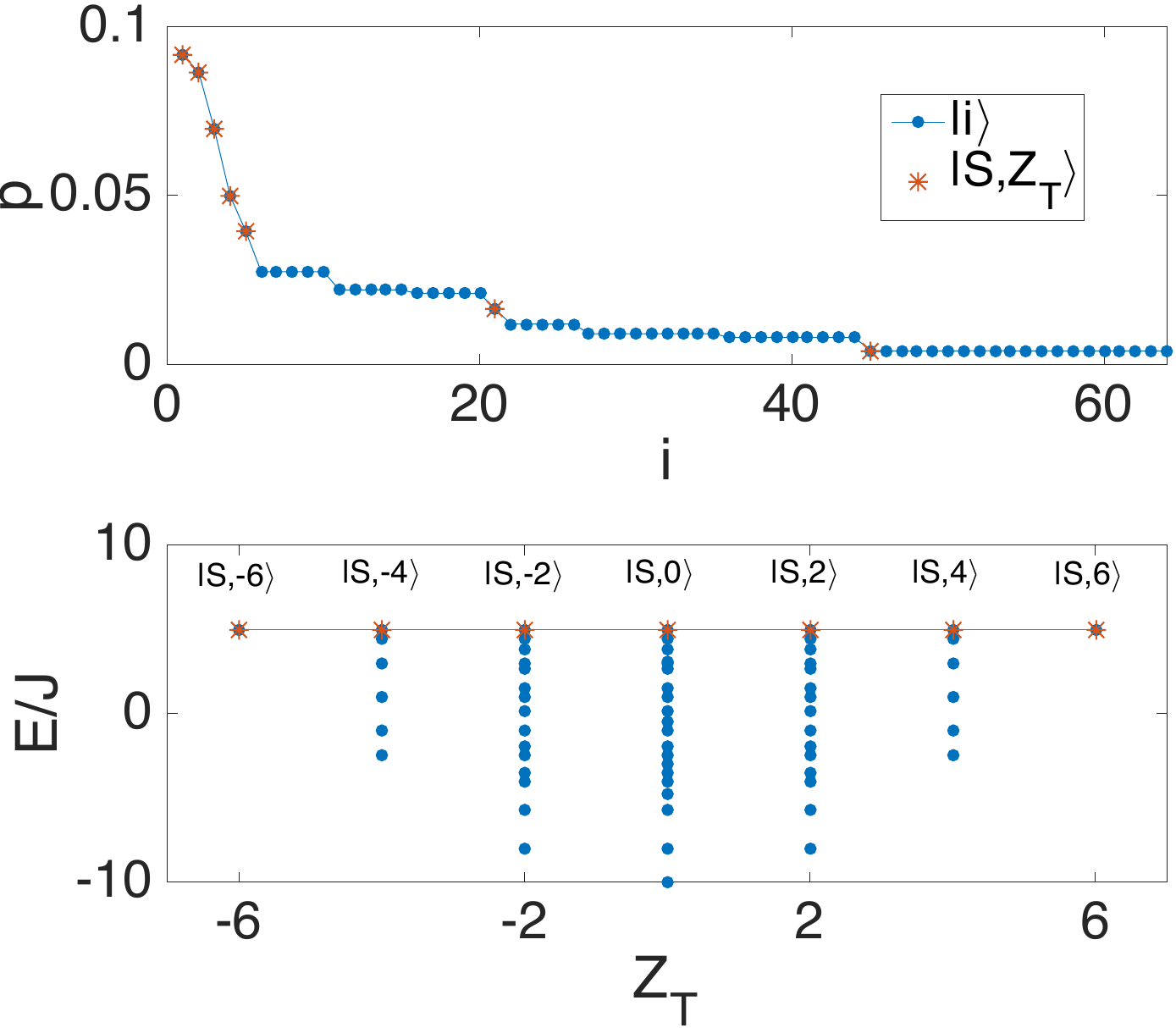}}\\ 
\caption{ 
(Top) The probabilities of all the $H_{XXZ}$ eigenstates, as computed from $\rho_{\mathrm{NESS}}$ for $L=6$ and $J=U$. (Bottom) The corresponding eigenenergies as a function of their magnetization. Highlighted are the $|S,Z_T\rangle$ states. The system parameters are $g = 0.1J$, $\kappa = 0.5g$, and $P=J$.
}
\label{EqNonEqU1}
\end{figure}

To further support this picture, in Fig.~(\ref{EqNonEqU1}) we provide the full spectral decomposition at the Heisenberg point, for the case of six emitters $L=6$. In the top panel, the probability of each state is plotted, and the bright states are highlighted in red. Again, it is clear that most of these states represent the most probable states. The bottom panel shows the state energy as a function of their magnetization, again with the bright states highlighted in red. The plot shows again the onset of resonance and highlights the average zero magnetization of the chain, as expected in a laser. 

\begin{figure}
{\includegraphics[width = 8.5cm]{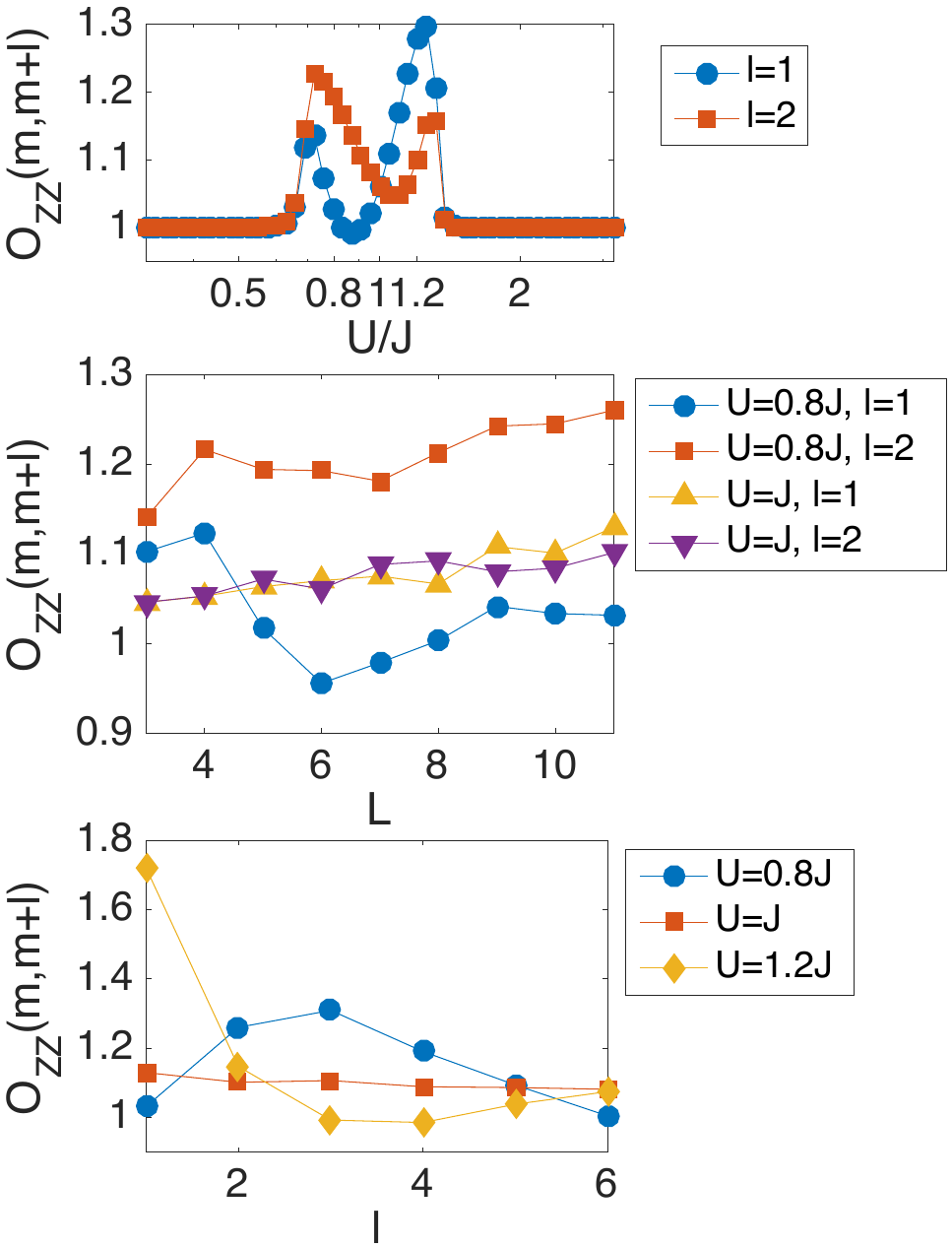}}\\ 
\caption{ (Top) Correlations between nearest-neighbors and next-nearest-neighbors with respect to the site $m=\mathrm{floor}(L/2)$ for a system of 5 emitters as a function of the interactions. (Middle) The correlations obtained from the stochastic Schr\"{o}dinger equation, for specific values of the ratio between interaction and hopping strength ($U=0.8J$ and $U=J$) as a function of the number of emitters, $L$. (Bottom) Spatial correlations obtained from the solution of the stochastic Schr\"{o}dinger equation for a chain of 11 sites as a function of the distance between emitters. The system parameters of these simulations are $g = 0.1J$, $\kappa = 0.5g$, and $P=J$.
}
\label{ExactScaling}
\end{figure}

\subsection{Correlations}

The most interesting features of the resulting manybody state of the emitters may be described by the correlations in the matter medium. In Fig.~(\ref{ExactScaling}) we show the charge correlations
\begin{equation}
O_{ZZ}(m,m+l)=\frac{\langle Z_mZ_l\rangle}{\langle Z_m\rangle \langle Z_l\rangle} \, ,
\end{equation} 
which allow to identify the type of order that is established throughout the chain. When either the hopping or the interactions dominate, there are no charge correlations since the emitters become fully inverted by their respective incoherent pumping rates. As the system enters the laser regime, spin correlations emerge as it is evidenced in the upper panel of Fig.~(\ref{ExactScaling}). For interaction values below the Heisenberg point ($U/J < 1$) the system shows a behavior reminiscent of a CDW, as indicated by the fact that the correlations between next-nearest-neighbors are larger than the ones between nearest-neighbors. 
At the Heisenberg point these two correlations become equal (up to statistical erros in the SSE simulations). Above the Heisenberg point ($U/J >1$), the correlations between nearest-neighbors become larger and the CDW order is lost.
Even though the system exhibits strong finite size effects, for chains up to 11 emitters the general picture described above is robust, and it is confirmed by our simulations for an increasing number of emitters shown in the middle panel of Fig.~(\ref{ExactScaling}). 
Precisely at the Heisenberg the all pair-correlations become equal irrespective of their distance (show in the lower panel of Fig.~(\ref{ExactScaling})). This further confirms that under such conditions the system behaves as if there was no direct couplings between the emitters, and the global field generates such long-range correlations. Also in the lower panel of Fig.~(\ref{ExactScaling})) we can see the charge-density behaviour emerging with a non trivial spacial period below the Heisenberg point for $U=0.8J$, while the correlations decay fast above the Heisenberg point for $U=1.2J$.

\section{Conclusions}

In summary, we have shown how the couplings of a many-body medium affects the properties of a laser. The many-body interactions strongly alter the emission and cooperativity of the atomic system which is consequently imprinted in the radiation emitted from the cavity. 
These results allow for the conclusion that in this model there exists an interplay between hopping and correlations that leads to the laser regime in a dynamical equilibrium reached as the balance between driving and dissipation. We have shown how the combination of hopping and interactions in the same proportion leads to a highly symmetric emission process at maximal efficiency.
We have also presented in detail the nonequilibrium state of the emitters and their resulting correlations.
The system exhibits interesting charge correlations that depend sensibly on the hopping-interaction ratio. In the absence of many-body interactions, one would expect only cavity-induced correlations of the ferromagnetic kind. Therefore, the CDW correlations in our case are a specific signature of the many-body nature of the system and in particular reflect the nature of the bright states involved in the lasing mechanism.
In remarkable contrast to the equilibrium behavior of the atomic system we show that the atomic correlations emerge as a genuine nonequilbrium phenomenon without an equilibrium analog which is due to the fact that the laser explores highly excited states of the chain of emitters.

\end{document}